\documentclass[rnote,oldversion]{aa}

\usepackage{graphicx}
\usepackage{txfonts}

\begin{document}

\title{Constraining the environment of GRB 990712 through emission line 
fluxes \thanks{Based on observations collected at the European Southern Observatory, Chile (ESO Programme 075.D-0771(A)).}}

\author{Ayb\"uke K\"upc\"u Yolda\c{s}\inst{1}
	\and Jochen Greiner\inst{1}
	\and Rosalba Perna\inst{2}}

\offprints{A. K\"upc\"u Yolda\c{s}, \email{ayoldas@mpe.mpg.de}}

\institute{
      Max-Planck-Institut f\"ur extraterrestrische Physik, 
              Giessenbachstrasse 1, D-85748 Garching, Germany
      \and JILA and Department of Astrophysical and Planetary Sciences, 
University of Colorado, 440 UCB, Boulder, 80309, USA }

\date{Received / Accepted }

\abstract
{The energy output in the gamma-ray burst (GRB) prompt
emission and afterglow phase is expected to photoionize the
surrounding medium out to large radii. Cooling of this gas produces
line emission, particularly strong in the optical, whose variability
is a strong diagnostics of the gas density and geometry in the close
environment of the burst.
We present the results of a spectral time
series analysis of the host galaxy of GRB~990712 observed up to
$\sim$6 years after the burst. We analyze the emission line fluxes
together with those of the previous observations of the same GRB, in
search for photoionization signatures.
We find that the emission line
fluxes show no variation within the uncertainities up to 6 years after
the burst, and we use the measured line intensities to set a limit on
the density of the gas within a few parsecs of the burst location.
This is the first time that emission from cooling GRB remnants is
probed on years time scales. 
}

\titlerunning{Constraining the environment of GRB 990712}
\maketitle

\section{Introduction}


In the fireball model of GRBs (e.g. Rees \& Meszaros \cite{rm92}), the
energy released from the collapse of a massive star is converted into
kinetic energy of thin baryonic shells which expand at
ultra-relativistic speeds. After producing the
prompt $\gamma$-ray emission by internal shocks between different
shells, the residual impacts on the surrounding gas and drives an
ultra-relativistic shock into the ambient medium. The shock accelerates 
relativistic electrons leading to the observed X-ray to radio 
afterglow radiation through synchrotron emission. The emission of the 
X-ray afterglow, integrated over the first 7--10\,days, typically contains 
the same energy as the primary $\gamma$-ray burst itself (Rees 
\& Meszaros \cite{rm98}).

Whatever the density of the GRB environment, 
photoionization of the circumburst material by the prompt
X-ray emission and by the X-ray and UV afterglow emission
will be inevitable. If the circumburst density is high, it will lead to 
time dependent (on hour timescale)
absorption (Perna \& Loeb \cite{pl}) and emission line features
(B\"ottcher et al. \cite{boett}; Ghisellini et al. \cite{ghis}),
such as those claimed to be seen in X-rays. On longer
timescales, the GRB photoionization may lead to indicative
recombination line features, which allow the identification of
remnants of GRBs in nearby galaxies (Band \& Hartmann \cite{band}; 
Perna, Raymond \& Loeb \cite{prl}, PRL in the following).

Photoionization of the ambient medium is well studied in the case of
supernovae which are the most similar cases to GRBs. For SN 1987A, IUE
observations showed that the prominent UV lines started to increase
simultaneously after 60--80\,days and stayed at a constant level until
400\,days after the initial exciting supernova outburst (Fransson et
al. \cite{fran89}).  After 400\,days most lines decreased fastly, and
reached the noise level by day 1500 (Sonneborn et
al. \cite{sonn}). Detailed modelling of the ionization zones and and the
line emission of the circumstellar gas (Lundqvist \& Fransson \cite{lf96}) 
allowed to constrain the gas density. 

In the case of a GRB, ionization by the prompt emission, the afterglow
(photon field) and by the blast wave (shock-ionization) will appear
additionally to the SN component, and largely dominate.  The blast wave
is expected to influence the ionization state of the gas on timescales
of hundreds to thousands of years after the burst (PRL). Therefore,
for the purpose of our analysis, we can safely assume that
photoionization is the dominating ionization mechanism. A large
fraction of the energy (e.g. kinetic, magnetic) stored in the GRB-jet
($\sim$10$^{52}$\,erg) is released in the afterglow.  The X-ray/UV 
component of this radiation
(with some contribution from the X-ray emission of the prompt phase)
is the main responsible for the ionization of the ambient medium. 
For typical GRB/afterglow luminosities, the size of the ionized region
is on the order of 100 pc for an ISM density of $\sim 1$ cm$^{-3}$ (PRL). 

After being ionized, the gas then starts to cool, on a timescale $t_{\rm
cool}\sim 10^5 (T/10^5\,{\rm K})/(n/{\rm cm}^{-3})$ yr, where $T$ is the
temperature of the gas, and $n$ the electron density.  Although
cooling starts soon after the burst of radiation has passed through
the medium, emission from the cooling gas increases during the early times
after the GRB. If a region of radius $R$ is heated and ionized by the
burst radiation, the maximum emission will
occur after a time $t\sim R/c\sim 3\;{\rm yr}\; (R/{\rm pc})$, due to 
light travel times from different parts of the region. 

The strength and timescale of the recombination emission depends strongly 
on the ambient density. While the modelling of the broad-band SED of 
afterglows has led to densities in the range 1-10 cm$^{-3}$, there are also 
observational indications for much higher densities: (i) observed variable 
X-ray lines (Watson et al. \cite{wat}; Reeves et al. \cite{reev}; Frontera 
et al. \cite{front}) and
continuum absorption (Lazzati \& Perna \cite{lp02}) require densities of 
$\sim$10$^5$-10$^6$ cm$^{-3}$;
(ii) some GRB afterglow data require a dense ($\sim$10$^4$ cm$^{-3}$) shell 
around some nearby low-density medium (1-10 cm$^{-3}$) 
(Chevalier et al. \cite{chev}).
The SN-GRB connection is now clearly proven for four GRBs 
(Galama et al. \cite{gala}; Hjorth et al. \cite{hjor}; 
Stanek et al. \cite{stan}; Malesani et al. \cite{male}; Pian et al. 
\cite{pian}), indicating 
the link between long-duration GRBs and deaths of massive stars. 
Observations of massive, Wolf-Rayet (WR) like stars, have shown 
that they loose matter via strong stellar winds. A WR stellar 
wind, interacting with a circumstellar medium,  
leads to the formation of a shell (termination shock) whose density and radial extent
depend on both the progenitor characteristics (i.e. mass loss rate,
wind velocity) as well as the density of the medium (e.g. Fryer et al. 
\cite{fry}). 

An ionized shell of, say, density $\sim 10^{3}$ cm$^{-3}$ and radial extent of a
few parsecs, will reach its peak emission on a timescale of a few years (due to
the light travel times), and will cool on a timescale of tens of years. 
Therefore, the first few years after the burst are crucial for detecting 
cooling emission from these dense, compact shells produced
by the wind termination shocks of the massive stars progenitors of (long) GRBs. 

This {\em Letter} reports a long-term monitoring, up to 6 years, of several strong
emission lines, and in particular the OIII$\lambda$5007 and the H$\beta$ lines, whose
ratio is an important discriminant of strongly photoionized, cooling gas 
as expected in GRB remnants (PRL). Our observations are described in \S 2, 
and the results are used in  \S 3 to set constraints on the environment of 
GRB 990712. Further implications for studies of galaxy properties are 
discussed in \S 4, while \S 5 summarizes our findings. 

\section{GRB 990712}

GRB 990712 was discovered by GRBM and WFC onboard BeppoSAX on July 
12.69655 UT, 1999. The duration of the burst was 30s and it was first 
localized by WFC at R.A. = 22:31:50, Dec. = -73:24.4 with an error radius of 
2$\arcmin$ (Heise et al. \cite{heis}). Follow-up observations led to the 
discovery of the GRB afterglow. The redshift of the burst is z=0.433 
(Vreeswijk et al. \cite{vree}). Two different groups found evidence for a 
SN bump from the optical lightcurve of the afterglow (Bj\"{o}rnsson et al. 
\cite{bjor}; Zeh et al. \cite{zeh}).
The BeppoSAX spectra of the prompt emission indicated a temporal emission 
feature located around 4.5 keV, which can be 
fit either with a Gaussian profile with rest frame energy of around 6.4 
keV that is consistent with an iron line, or with a blackbody spectrum 
with kT $\sim$ 1.3 keV 
(Frontera et al. \cite{front01}). Since the iron line 
interpretation requires a very high density environment that would obscure 
the afterglow, Frontera et al. (\cite{front01}) prefer 
the thermal component interpretation which can be accounted for by the 
fireball model.

\begin{table}
\caption{Log of observations}
\label{tab:log}
\begin{tabular}{ccccc}
\hline
\hline
 Date & Instrument & Grism & Coverage & Exposure time\\
 & & & (nm) & (sec) \\
\hline
14 Jul 1999 & FORS1 & 150I & 370 -- 770 & 2400\\
11 Nov 1999 & EFOSC & Gr6 & 400 -- 800 & 1800\\
06 Jun 2002 & FORS1 & 600R & 525 -- 745 & 4320\\
17 Jul 2004 & FORS2 & 600RI & 512 -- 845 & 2400\\
05-06 Jul 2005 & FORS2 & 300V & 445 -- 850 & 7200\\
\hline
\end{tabular}
\end{table}

The host galaxy of GRB 990712 is one of the 
brightest GRB host galaxies with V = 22.3 mag and R = 21.8 mag (Sahu 
et al. \cite{sahu}).
We obtained spectra of GRB 990712 on July 05 and 06, 2005, 
approximately 6 years after the GRB using VLT/FORS2 under 
good seeing conditions ($\sim$0$\farcs$6). The 300V/GG435 
grism/filter was used with a slit width of 1$\farcs$0. The total exposure 
time was 2 hours (4$\times$30 minutes). The spectra were reduced using 
standard IRAF routines and 
calibrated using standard star G158-100 observed on July 05, 2005 with the 
same grism/filter. The flux calibration was further validated by folding 
the spectrum with the FORS R-band filter curve and comparing the obtained 
magnitude with that of the host galaxy given in Christensen et al. 
(\cite{chris}). We corrected the spectra for foreground 
extinction of E(B -- V) = 0.03 (Schlegel et al. \cite{schle}). The 
line fluxes were determined by fitting a Gaussian to the line using the 
SPLOT task of IRAF. The continuum level was determined locally.

\begin{figure}
\centering
\includegraphics[height=6cm,angle=0]{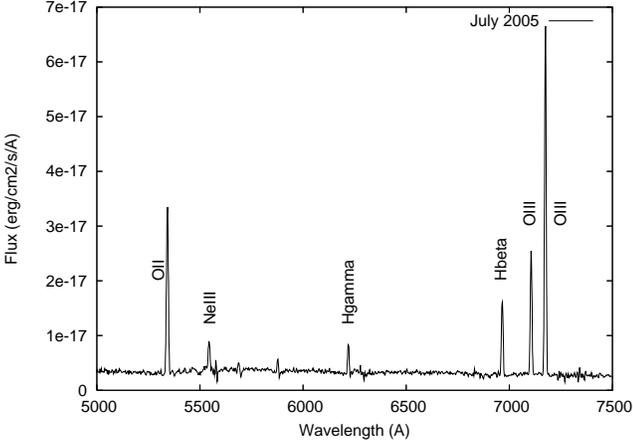}
\caption{Spectra of the host galaxy of GRB 990712 obtained in July 2005.}
\label{fig:spec1}
\end{figure}

\begin{figure}
\centering
\includegraphics[height=6cm,angle=0]{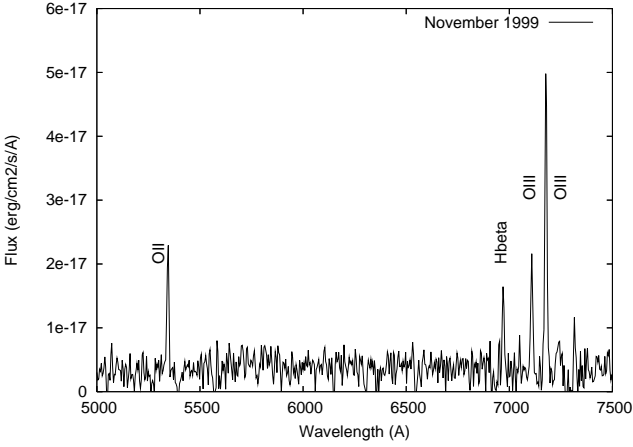}
\caption{Spectra of the host galaxy of GRB 990712 obtained in November 1999.}
\label{fig:spec2}
\end{figure}

The line fluxes are compared with those derived using VLT archival
data of observations obtained on July 14, 1999 (PI: Galama), November
11, 1999 (PI: Courbin), June 06, 2002 (PI: Mirabel) and on July 17,
2004 (PI: Le Floc'h) (see Tab.\ref{tab:log}). The spectra were treated in 
an identical way, and calibrated using standard stars EG 274 (Sep 15, 
1999), LTT 377 (at Dec 6, 1999 and at Jul 15, 2004), and LTT 3854 (Jun 05, 
2002) for July 1999, November 1999, July 2004 and June 2002 data, 
respectively. July 1999 data was further flux calibrated using the 
afterglow brightness at that time taken from Sahu et   
al. (\cite{sahu}). The flux values we obtain are consistent with the
published values obtained by flux calibrating the same data using the
afterglow brightness extrapolated in time (Vreeswijk et
al. \cite{vree}). The previously unpublished 2002 and 2004 fluxes were 
similarly validated as in the case of 2005 spectra. The estimated error 
in the flux calibration is about 10$\%$ for all data. The line fluxes are 
shown in Table \ref{tab:flux}. The flux errors in Table \ref{tab:flux} 
only include the uncertainities in the continuum level. 
Figure \ref{fig:spec1} shows our spectrum from July 
2005 and Figure \ref{fig:spec2} shows the spectrum from November 1999 which has the worst signal-to-noise ratio. 

\begin{table*}
\centering
\caption{Emission line fluxes}
\label{tab:flux}
\begin{tabular}{cccccc}
\hline
\hline
 Line & \multicolumn{5}{c}{Flux (10$^{-16}$ erg/s/cm$^{-2}$)}\\
\hline
 & at day 1.5& at day 123& at day 1060& at day 1832& at day 2185\\
\hline
[O II]($\lambda$3727)&3.52$\pm{0.15}$ &  &3.66$\pm{0.20}$ & &3.40$\pm{0.60}$ \\
${\rm [Ne III]}$($\lambda$3869) &0.5$\pm{0.1}$ &  &0.55$\pm{0.07}$ 
&0.59$\pm{0.05}$ &0.50$\pm{0.05}$ \\
H$\gamma$($\lambda$4340) &0.3$\pm{0.1}$ & &0.53$\pm{0.12}$ &0.46$\pm{0.05}$ &0.46$\pm{0.05}$ \\
H$\beta$($\lambda$4861) &1.15$\pm{0.15}$ &1.30$\pm{0.15}$ &1.29$\pm{0.09}$ &1.34$\pm{0.04}$ &1.33$\pm{0.05}$ \\
${\rm [O III]}$($\lambda$4959) &2.25$\pm{0.15}$ &2.33$\pm{0.18}$ 
&2.10$\pm{0.15}$ &2.18$\pm{0.08}$ &2.22$\pm{0.08}$ \\
${\rm [O III]}$($\lambda$5007) &6.15$\pm{0.15}$ &6.17$\pm{0.25}$ 
&6.00$\pm{0.10}$ &5.97$\pm{0.08}$ &6.08$\pm{0.08}$ \\
\hline
\end{tabular}
\end{table*}

\begin{figure}
\includegraphics[height=6cm,angle=0]{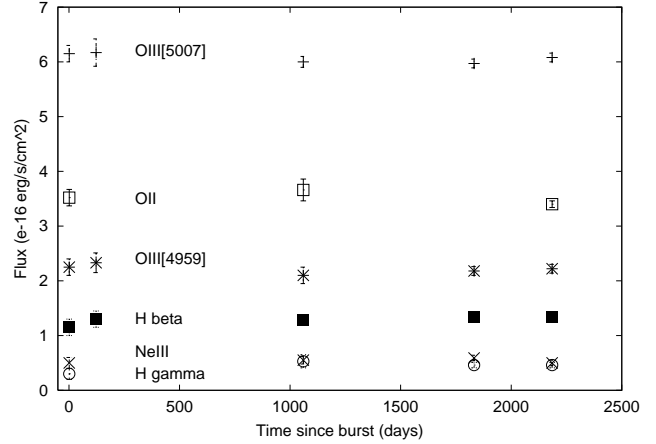}
  \caption{ [Ne III], H$\gamma$, H$\beta$,
[O II] and [O III] 4959, 5007 line flux light curves for GRB 990712. Error bars 
are smaller than the symbol size, if not visible.}
\label{fig:lc}
\end{figure}

All of the observed lines, which are [O II], [Ne III], H$\gamma$, H$\beta$
and [O III] ($\lambda\lambda$4959,5007) lines, are observed to have constant fluxes over $\sim$6
years after the burst (see Fig.\ref{fig:lc}). To overcome the
difficulty of comparing line fluxes from different spectra obtained
with different settings, i.e. instrument, grism, night conditions
etc., we derive our results on the circumburst environment based
particularly on the [O III]$\lambda$5007 and the H$\beta$ lines. These are
close enough in wavelength space to overcome the possible effects of
different flux calibrations, and at the same time, as discussed below,
they are especially strong discriminants of photoionization vs. collisional
ionization.

\section{The circumburst environment}

A distinctive signature of photoionized gas is a high [O III]
$\lambda$5007/H$\beta$ ratio ($\ga 5$), which is not generally produced in steady shocks
for solar abundances (Shull \& McKee \cite{shull}).  This ratio increases 
with the ionization
parameter and, for typical X-ray/UV photoionizing fluxes of GRBs and
their afterglows, it reaches a value on the order of 100 (PRL). Our
observations at about 6 years after the burst, as well as the previous
ones at earlier times, show that [O III] $\lambda$5007/H$\beta\sim 4.6$ in the
host of GRB 990712. Combined with the log ([O II]/H$\beta$) value of 
0.4$\pm{0.1}$, this value is rather typical of an HII galaxy, as 
can be seen by comparison with the sample of
emission line galaxies at redshifts $0\la z\la 0.3$ from the
Canada-France redshift survey (Rola, Terlevich \& Terlevich \cite{rola};
see also Vreeswijk et al. \cite{vree}).  
Therefore, the constraints that we are
able to put on the close environment of GRB 990712 will derive from
the lack of contribution from photoionized gas to the brightest
expected line ([O III] $\lambda$5007) from the cooling gas, within the
observational uncertainties.  The measured flux of this line is $\sim
6\times 10^{-16}$ erg/s/cm$^{-2}$ to within $\sim 10\%$ uncertainty
(see Table 2). At a redshift of 0.433, this flux corresponds to a
luminosity of $\sim 3.8\times 10^{41}$ erg/s for a LCDM cosmology with
Hubble parameter h=0.73 (Spergel et al. \cite{spe}).  Since the luminosity
is constant to within the 10\% uncertainty value of $\sim 3\times 10^{40}$ erg/s,
the contribution from the cooling GRB remnant cannot be larger than
this value at any given time during the observation window.

Numerical simulations of cooling GRB remnants (PRL) show that the
radiation flux from a typical GRB and its afterglow ionizes a region
on the order of $\sim 100$ pc for an ISM density of $\sim 1$
cm$^{-3}$. For a gas of solar metallicity, the corresponding
luminosity of the [O III] $\lambda$5007 line from the cooling gas is found to reach
a value of $\sim 10^{38}$ erg s$^{-1}$ over a time $t_{\rm cool}\sim$ a few $\times 10^4$ yr
(see Fig.3 in PRL; the details of the computation of the cooling
radiation can also be found in PRL).  The brightness of the line
scales with $n\,n_e$, where $n_e$ is the electron density. For a gas
metallicity not too far from solar (so that the particle number is
dominated by H) and a highly ionized gas, one has $n_e\sim n$, and the
line luminosity can be scaled, to a first approximation,
as\footnote{Note however that, as the density increases, the maximum
size of the region that can be ionized will
decrease. Furthermore, note that the details of the rise time also depend on the extent
of the beaming of the ionizing radiation, but we do not worry about 
secondary  effects here since the data only allow us to set upper limits.} 
$\sim 10^{38} (n/{\rm cm}^{-3})^2\;(R_e/100\,{\rm pc})^3$ erg
s$^{-1}$.  Our observations, up to 6 years after the burst, allow us
to probe an emitting region, $R_e$, of at most 2 pc in size, due to
light delay effects as discussed in \S 1.  Since no flux variation has
been observed within the 10\% flux error (corresponding to a luminosity of
about $3\times10^{40} {\rm erg}\;{\rm s}^{-1}$), we deduce that the
line luminosity due to the cooling gas has to be below this level,
i.e.
\begin{equation}
L_{5007} \sim 10^{38}
(n/{\rm cm}^{-3})^2\;(R_e/100\;{\rm pc})^3 {\rm erg}\;{\rm s}^{-1}\la
3\times10^{40} {\rm erg}\;{\rm s}^{-1}\;
\end{equation}
which yields the limit on the density $n\la 6\times 10^3$ cm$^{-3}$
for $R_e=2$ pc. 
This limit can be used to constrain the range of allowed
parameters for the GRB host and ISM densitites. For
example, Fryer et al. (\cite{fry}) find that, for a progenitor 
star with mass loss $\dot{M}=10^{-5}M_\odot$ yr$^{-1}$ and
wind velocity $\sim 1000$ km/s, expanding in a medium of density
in the range $\sim 10^3-10^4$ cm$^{-3}$, the inner radius of the
shell is on the order of tenths of a parsec and the outer radius
is $>2$ pc (with the shell density being on the order of the ISM density). 
These large shell densities, filling a region up to the observed
emitting volume of $\sim 2$ pc, are not favoured by our observations,
since they would likely result in a variable [O III] $\lambda$5007 flux over
the 6 years of observations. On the other
hand, termination shock shells produced in the impact of the 
wind with a lower density medium are consistent with the
lack of variability. Also note that, given the 6 year timescale
of the observations, termination shocks located at distances $\ga 2$ pc
cannot be ruled out by the currently available data.  Generally speaking,
observations at longer timescales allow one to probe shell terminations shocks
over a larger range of distances from the GRB progenitor star. However,
since more distant termination shocks are generally associated with
lower ISM densitites (and hence a lower luminosity of the [O III] $\lambda$5007 line),
in order to separate the eventual contribution of this line due
to the cooling gas in the close GRB environment, from that due to the host 
galaxy itself, a higher signal-to-noise ratio in the observations as well 
as a consistent set of observations (i.e. observations obtained with the 
same instrument and settings) is necessary. 

\section{Star formation rate and metallicity}

The star formation rate (SFR) of the host galaxy of GRB 990712 has been
previously calculated based on line fluxes and ultra-violet flux, and also an upper limit was derived based on radio non-detection (Hjorth et al. 
\cite{hjor00}; Vreeswijk et al. \cite{vree,vreeb}; 
Christensen et al. \cite{chris}). 
Using the line fluxes of our July 2005 spectrum and the same method used 
by Vreeswijk et al. (\cite{vree}), we calculated SFR$_{OII}$ = 
2.8$^{+0.4}_{-0.9}$ M$_{\sun}$/yr and A$_{V}$ = 1.7$^{+0.9}_{-0.8}$ mag 
based on the H$\gamma$/H$\beta$ ratio. Our A$_{V}$ value is on average 
lower than the A$_{V}$ = 3.4$^{+2.4}_{-1.7}$ derived by Vreeswijk et al. 
(\cite{vree}), therefore our extinction corrected SFR value (SFR$_{OII 
corr}$ = 10$^{+15}_{-6}$ M$_{\sun}$/yr) is also lower than their 
calculation. Nonetheless both A$_{V}$ and the extinction 
corrected SFR are in agreement with Vreeswijk et al. (\cite{vree,vreeb}) 
values within the errors. 

We do not detect the [O III] $\lambda$4363 line which is
necessary for the determination of the electron temperature T$_{e}$ by
means of lines (like [O III], [Ne III] etc.) from high ionization zone
elements. Therefore, we cannot estimate the oxygen and neon abundance
based on the electron temperature. However we can still estimate the
oxygen abundance based on the ratio R$_{32} = ({\rm [O II]}
\lambda3727 + {\rm [O III]} \lambda\lambda4959,5007 )/
H_\beta$. Kewley \& Dopita (\cite{kp02}) suggest to use the relation
given by Zaritsky et al. (\cite{z94}) to obtain an estimate of the
oxygen abundance. Using the formulae given in Zaritsky et
al. (\cite{z94}) we obtain log(O/H)= -3.7$\pm{0.1}$. However, that
formula is calibrated for metal rich galaxies and overestimates the
metallicity for values log(O/H) $<$ -3.5 (see Kobulnicky et
al. \cite{kob99}, Kewley \& Dopita \cite{kp02}). Therefore we used
equation (16) of Kobulnicky \& Kewley (\cite{kk04}), which is
adopted from the relation given by Kewley \& Dopita (\cite{kp02}) and
parameterized for the lower metallicity branch (log(O/H) $<$
-3.6). The result is log(O/H)= -3.7$\pm{0.1}$, which is the same as
our initial estimate using the relation given by Zaritsky et
al. (\cite{z94}).  Similarly, Vreeswijk et al. (\cite{vree}) obtained
-3.7$\pm{0.4}$ for log (O/H), which is in agreement with our
estimate. The oxygen abundance we obtained is just a bit lower than the
solar value (log(O/H) = -3.34; e.g. Asplund et al. 2005).
 Therefore our
assumption that the metallicity of the gas nearby the GRB is not far
from solar is reasonable in our derivation of the [O III] line
luminosity of the cooling gas.

It should be noted how, in the context of GRBs, studying the
photoionization signatures of cooling gas not only helps understand
the nature of the GRB progenitor star, but it also helps reduce
possible biases in the determinations of two important properties
of the GRB host galaxy such as the SFR and the metallicity.  In fact,
SFR and metallicity calculations based on emission lines generally
rely on observations obtained at a single epoch.  However, in order to
properly assess the possible level of contamination by the cooling GRB
remnant, multiple epochs of observations spanned over a long timescale
are necessary.

\section{Summary}

We have presented the results of the spectral analysis of the host galaxy
of GRB 990712. With the last set of observations taken about 6 years after the burst,
this is the longest time coverage for a GRB host galaxy up to date. 
Though we do not detect line variations, timescales of a few years are
important for detecting cooling radiation from the heated shells produced
by the wind termination shocks of the massive star progenitors of the GRBs. 
For the case of the GRB 990712 host, the lack of time variability in the
[O III] $\lambda$5007 line, combined with the $\la 5$ ratio of the [O III]$\lambda$5007/H$\beta$ 
lines, has allowed us to set an upper limit to the contribution from the
cooling gas. This limit, in turn, could be used to constrain the allowed
range of densitites within a region of about 2 pc surrounding the burst.
We have therefore shown how this type of observations provides a useful
complement to the studies of the close environment of GRB progenitors and,
therefore, can help reconstruct the characteristics of the
GRB progenitor star.

Finally, we have pointed out how, if a substantial contamination to 
the galaxy spectra is provided by the GRB cooling radiation, inferences
of the SFR and metallicity that are drawn from measurements of line ratios
can be biased. To be able to assess the degree of this contamination, long-term
monitoring of the GRB host galaxies is necessary. 

\begin{acknowledgements}
   We thank Arne Rau and Markus B\"{o}ttcher for insightful comments 
and discussions that contributed to the early stages of this work, and 
John Raymond for very useful comments on the manuscript.
   AKY acknowledges support from the International Max-Planck Research 
   School (IMPRS) on Astrophysics. 
RP acknowledges support from NASA under grant NNG05GH55G, and
from the NSF under grant AST~0507571.
\end{acknowledgements}

\end{document}